\documentclass{article}
\usepackage{amsmath, amsthm, amssymb,amsbsy,bbm}
\usepackage{hyperref} 
\hypersetup{%backref,  %pdfpagemode=FullScreen, 
colorlinks=false, linkcolor=blue, urlcolor=blue, citecolor=red}
\def\bea{\begin{eqnarray}}
\def\eea{\end{eqnarray}}
\def\bean{\begin{eqnarray*}}
\def\eean{\end{eqnarray*}}

\def\bvec#1{\raise1.5ex\hbox{$\rightarrow$}\mkern-16.5mu #1}

\def\m#1{\mathcal#1}

\newcommand{\be}{\begin{equation}}
\newcommand{\ee}{\end{equation}}

\newcommand{\barr}{\begin{array}}
\newcommand{\earr}{\end{array}}

\newcommand{\bed}{\begin{displaymath}}
\newcommand{\eed}{\end{displaymath}}
\newcommand{\bal}{\begin{array}{ll}}
\newcommand{\eal}{\end{array}}

\renewcommand{\d}{\partial}
\newcommand{\nn}{\nonumber}
\newcommand{\ep}{\epsilon}
\newcommand{\Ese}{E_{7(7)}}
\newcommand{\Eei}{E_{8(8)}}
\newcommand{\E}{E_{7(7)}}
\newcommand{\ka}{\kappa}

\numberwithin{equation}{section} 
\begin{document}
\begin{titlepage}
\begin{flushright}    UFIFT-HEP-08-05\\ 
%hep-ph/yymmnnn 
\end{flushright}
\vskip 1cm
\centerline{\LARGE{\bf {$E_{8(8)}$  in Light Cone Superspace}}} 
\vskip 2cm
\centerline{\bf Lars Brink} %${}^{\,a}_{}$, Sung-Soo Kim${}^{\,b}_{}$, and Pierre Ramond${}^{\,b}_{}$}
\vskip .2cm
\centerline{\em %${}^{a~}_{}$
Department of Fundamental Physics}
\centerline{\em Chalmers University
of Technology, }
\centerline{\em S-412 96 G\"oteborg, Sweden}

\vskip .2cm
\centerline{\bf  Sung-Soo Kim, % ${}^{\,b}_{}$
 and Pierre Ramond  %${}^{\,b}_{}$
}
\vskip .2cm
\centerline{\em  %$${}^{b~}_{}$
Institute for Fundamental Theory,}
\centerline{\em Department of Physics, University of Florida}
\centerline{\em Gainesville FL 32611, USA}
\vskip 2cm
 
\ \centerline{\bf {Abstract}}
\vskip .5cm
\noindent 
\noindent We derive the non-linear action of $E_{8(8)}$ on the constrained chiral superfield in the light-cone superspace with eight complex Grassmann variables. We construct (to lowest order in the coupling) the sixteen dynamical supersymmetries which generate a Hamiltonian with $\Eei$  invariance in three space-time dimensions, and show that it  has only interactions with even powers of the coupling constant.    
\vfill
\begin{flushleft}
April 2008 \\
\end{flushleft}
\end{titlepage}
\section{Introduction}
The degrees of freedom of maximally supersymmetric theories in various dimensions are represented in light-cone superspace by a constrained chiral superfield, whose components represent the $256$ fields of several theories: $\m N=1$ supergravity in eleven dimensions \cite{Cremmer:1978km},  $\m N=8$ supergravity in four dimensions  \cite{deWit:1977fk, Cremmer:1979up, de Wit:1982ig}, and $\m N=16$ supergravity in three dimensions \cite{Julia:1982tm, Marcus:1983hb}. In a previous paper, we have shown how the Cremmer-Julia \cite{Cremmer:1979up} non-linearly realized $\E$ symmetry acts on this superfield \cite{Brink:2008qc}, and how it can be used to construct its interactions.

In this paper, we construct the non-linear $E_{8(8)}$ transformations on the same superfield. Dynamics is introduced by constructing  $16$ dynamical supersymmetries in three dimensions. In particular, $SO(16) (\subset E_{8(8)}$) invariance requires the dynamical supersymmetries to be limited to terms odd in the superfield: the $d=3$ $E_{8(8)}$-invariant  theory has no vertices of odd order (cubic, quintic, etc.).  This is understandable since the superfield contains the two $SO(16)$ spinor representations, and spinor representations have no odd invariants. Thus this theory is different from that obtained by dimensional reduction, which is {\em not} $E_{8(8)}$ invariant. 

%%%%%%%%%%%%%%%%%%%%%%%%%%%%%%%%%%%%%%%%%%%%%%%%%%%
\section{Chiral Superspace}
Consider the $N=8$ superspace spanned by eight Grassmann variables, $\theta^m$ and their complex conjugates $\bar\theta_m$ $(m=1,...,8)$. 
Introduce the chiral derivatives

\bea
d^m ~\equiv~ -\frac{\d}{\d\bar\theta_m} -\frac{i}{\sqrt{2}} \theta^m\d^+ \, , ~~~ 
\bar d_m ~\equiv~ \frac{\d}{\d\theta^m} +\frac{i}{\sqrt{2}} \bar\theta_m\d^+ \ , 
\eea
written in terms of the light-cone derivative, $\partial^+$, where  

\be
 ~ {\partial^{\pm}}=\frac{1}{\sqrt 2}\,(\,-\,{\partial_0}\,{\pm}\,{\partial_{d-1}}\,)\ ,
\ee
are conjugate to ${x^{\pm}}=\frac{1}{\sqrt 2}\,(\,{x^0}\,{\pm}\,{x^{d-1}}\,)$, with the metric $(-,+,\cdots,+)$ in the space with $(d-2)$ transverse coordinates $x^{\,j}_{\bot}$ ($j=1, \cdots, d-2$).
The chiral derivatives satisfy the canonical anticommutation relations

\begin{equation}
\left\{ d^m\,,\,\bar d_n\right\}~=~ -i \sqrt{2}\delta^m{}_{n} \d^+ \ .
\end{equation}
They are used to construct a constrained chiral superfield $\varphi$ and its complex conjugate $\overline\varphi$, related by the {\em inside-out constraint}

\begin{equation}
\varphi~=~\frac{1}{4\,\d^{+4}}\,d_{}^1d_{}^2\cdots d_{}^8\, \overline\varphi\ ,
\end{equation}
as well as the chiral constraints

$$
d_{}^m\, \varphi ~=~0\ ,\qquad \bar d^{}_m \,\overline \varphi ~=~0 \ .
$$
The chiral superfield can then be expanded in powers of $\theta^m$, 

\begin{eqnarray}\label{superfield}
\varphi\,(\,y\,)\,&=&\,\frac{1}{{\d^+}^2}\,h\,(y)\,+\,i\,\theta^m\,\frac{1}{{\d^+}^2}\,{\overline \psi}_m\,(y)\,+\,i\,\theta^{mn}_{}\,\frac{1}{\d^+}\,{\overline B}_{mn}\,(y)\nn\  \\
\;&&-\,\theta^{mnp}_{}\,\frac{1}{\d^+}\,{\overline \chi}^{}_{mnp}\,(y)\,-\,\theta^{mnpq}_{}\,{\overline D}^{}_{mnpq}\,(y)+\,i\widetilde\theta^{}_{~mnp}\,\chi^{mnp}\,(y)\nn \\
&&+\,i\widetilde\theta^{}_{~mn}\,\d^+\,B^{mn}\,(y)+\,\widetilde\theta^{}_{~m}\,\d^+\,\psi^m_{}\,(y)+\,{4}\,\widetilde\theta\,{\d^+}^2\,{\bar h}\,(y)\ ,
\end{eqnarray}
where 

\[
\theta^{a_1a_2...a_n}_{}~=~\frac{1}{n!}\,\theta^{a_1}\theta^{a_2}_{}\cdots\theta^{a_n}_{}\ ,\quad \widetilde\theta^{}_{~a_1a_2...a_{n}}~=~ \epsilon^{}_{a_1a_2...a_{n}b_1b_2...b_{(8-n)}}\,\theta_{}^{b_1b_2\cdots b_{(8-n)}}\,\ .
\]
The expansion coefficients are functions of the chiral coordinates 

$$
y~=~(x^{\,j}_{\bot},\, x^+, \,y^-\equiv x^- -\frac{i}{\sqrt2}\theta^m\bar\theta_m\, )\ ,
$$
and can be viewed as the $256$ physical fields of theories in various dimensions. In four dimensions (two transverse coordinates), they describe the physical degrees of freedom of $\m N =8$ Supergravity:  $128$ bosons, the spin-2 graviton $h$ and $\overline h$, twenty eight vector fields $\overline B_{mn}$ and $ B^{mn}$ and seventy real scalars ${\overline D}_{mnpq}$; $128$ fermions: eight spin-$\frac{3}{2}$ gravitinos ${\psi}^m$ and ${\overline \psi}_m$, fifty six  gauginos ${\overline \chi}_{mnp}$ and their conjugates $\chi^{mnp}$. In eleven dimensions, they encode the three fields of $\m N=1$ Supergravity \cite{Ananth:2005vg}. In three dimensions, it can be used to describe at least two different $\m N=16$ Supergravity theories with $128$ scalars and $128$ fermions, but they differ in their global non-linear symmetries, as we show in this paper.

%%%%%%%%%%%%%%%%%%%%%%%%%%%%%%%%%%%%%%%%%%%
\section{ Symmetries of $N=8$ Superspace}
In $N=8$ Superspace, we can also introduce the operators   

\begin{equation}
q^m ~=~ -\frac{\d}{\d\bar\theta_m} +\frac{i}{\sqrt{2}} \theta^m\d^+ \, , \qquad
\bar q_m ~=~ \frac{\d}{\d\theta^m} -\frac{i}{\sqrt{2}} \bar\theta_m\d^+ \ ,
\end{equation}
which satisfy the anticommutation relations

\begin{equation}
\{\, q^m\, , \, \bar q_n\,\} ~=~ i\sqrt{2}\, \delta^m{}_n \,\d^+\ .
\end{equation}
Their linear action on the chiral superfield 

\bea
\delta^{kin}_{\bar s}\,\varphi(y)~=~\overline \ep_m^{}\,q_{}^m\,\varphi(y)\ , \qquad \delta^{kin}_{s}\,\varphi(y)~=~ \ep^m\,\bar q_m\,\varphi(y)\ ,
\eea
where $\ep_m$ and $\overline \ep_m$ are Grassmann  parameters, do not alter  chirality, since 

\be
\{\, q^m\, , \, \bar d_n\,\} ~=~ \{\, q^m\, , \, d_n\,\} ~=~0\ .
\ee
These transformations are interpreted as the kinematical light-cone supersymmetries. 
\vskip .3cm

On the other hand, their quadratic action on the chiral superfields generates the $120$ $SO(16)$ transformations. The eight Grassmann variables and their conjugates form its vectorial $\bf 16$ representation under

$$
SO(16)~\supset~ SU(8)\,\times\, U(1)\ ,\qquad {\bf 16} ~=~ \bf8\,+\,\overline{\bf 8}\ .
$$
The $SU(8)$ and $U(1)$ generators are given by

\begin{equation}\label{SU8andU1}
T^i{}_j ~=~ \frac{i}{2\sqrt{2} \,\d^+} \left( q^i \bar q_j\, -\,\frac{1}{8}\,\delta^i{}_j\, q^k \bar q_k \right)\ ,\,\qquad T~=~\frac{i}{4 \sqrt{2} \, \d^+}\,[\, q^{k}_{}\,,\,\bar q_{k}\, ]\ ,
\end{equation}
with commutation relations

\begin{equation*}
 [\,T^i{}_j\,,\,T^k{}_l\,] ~=~\delta^k{}_j\, T^i{}_l - \delta^i{}_l \,T^k{}_j \ ,\qquad
[\,T\,,\,T^i{}_j\,]~=~0\ .
\end{equation*}
The remaining quadratic combinations describe the coset transformations $SO(16)/(SU(8)\times U(1))$

\begin{equation}\label{28}
T^{ij}~=~\frac{1}{2}\frac{1}{\d^+} q^iq^j\, , \qquad T_{ij}~=~\frac{1}{2}\frac{1}{\d^+} \bar q_i \bar q_j\ ,
\end{equation}
which form the ${\bf 28}$ and ${\bf \overline{28}}$ of $SU(8)$, and close on ($SU(8) \times U(1)$) 

\begin{equation*}
\ [\, T^{ij}\,,\, T_{kl}\,  ]~ = ~\delta^j{}_k T^i{}_l \,-\,
\delta^i{}_k T^j{}_l\, -\,\delta^j{}_l T^i{}_k  \,+\,\delta^i{}_l T^j{}_k \, +\, 2\,(\,\delta^j{}_k\delta^i{}_l \,-\,\delta^j{}_l \delta^i{}_k \,)\, T \ .
\end{equation*}
$SO(16)$ acts linearly on the chiral superfield

\[
 \delta^{}_{SU_8} \, \varphi~=~ \omega^j{}_{i}\,T^{i}{}_{j}\, \varphi\ ,~\quad \delta_{U(1)}\, \varphi ~=~ T \,\varphi\ ,
 \]
\begin{equation}
\delta_{\bf 28}\,\varphi~=~ \alpha_{ij}\,\frac{q^iq^j}{\d^+}\, \varphi\ , \qquad 
\delta_{\bf \overline{28}}\,\varphi ~=~ \alpha^{ij} \frac{\bar q_i \bar q_j}{\d^+}\, \varphi \ ,
\end{equation}
where $\omega^j{}_{i}$, $\alpha_{ij}$, and $\alpha^{ij}$  the transformation parameters.
$SO(16)$ is the largest {\it linearly} realized symmetry group in $N=8$ Superspace.

%%%%%%%%%%%%%%%%%%%%%%%%%%%%%%%%%%%%%%%%%%%%%%%%%%%%%%
\subsection{ $\Eei$ Symmetry} 
In this section, we show how $SO(16)$  can be extended to $\Eei$, the largest non-compact group that contains $SO(16)$ as its maximal compact subgroup. In a previous paper \cite{Brink:2008qc}, we had extended the $SU(8)$ symmetry of $N=8$ light-cone Superspace to the non-compact $\Ese$ with  $SU(8)$ as its maximal compact subgroup. While $SU(8)$ is linearly realized, the seventy coset $\Ese/SU(8)$ transformations act non-linearly on the chiral superfield in any dimensions. However in four dimensions, $\Ese$ commutes with the light-cone Hamiltonian, reproducing the well-known Cremmer-Julia {\it dynamical} symmetry \cite{Cremmer:1979up, de Wit:1982ig} of  $\m N=8$ Supergravity.

We showed \cite{Brink:2008qc} that the non-linear $\Ese/SU(8)$ coset transformations in $N=8$ Superspace of the chiral superfield could be elegantly expressed as   

\bea\label{70transf}
\delta_{\bf 70}\,\varphi&=&\delta_{\bf 70}^{(-1)}\,\varphi \,+\,\delta_{\bf 70}^{(1)}\,\varphi\,+\, \m O(\ka^3) \nn\\
&=&-\frac{1}{\kappa}\,\theta^{klmn}_{}\,\overline\beta^{}_{klmn}\nn\\
&&+\, \frac{\ka}{4\,\cdot\,4!}\,\beta^{mnpq}\left(\frac{\d}{\d\, \eta} \right)_{mnpq}\,\frac{1}{\d^{+2}} \left( e^{\eta\cdot \hat{\bar d} }  \,\d^{+3}\varphi\, e^{-\eta\cdot \hat{\bar d} } \d^{+3}\varphi\,\right)\bigg|_{\eta=0}\nn\\
&&+\, \m O(\ka^3)\ ,
\eea
order by order in the dimensionful parameter $\ka$, and where $\beta^{ijkl}$ are the seventy coset $\Ese/SU(8)$ parameters which satisfy the self-duality condition

$$\beta^{ijkl} ~=~\frac{1}{4!}\, \ep^{ijklmnpq}\,\overline \beta_{mnpq} \ , $$
$\hat{\bar d}_m\equiv{\bar d_m}/{\d^+}$, and $\eta^m$ are Grassmann variables with

$$ \left(\frac{\d}{\d\, \eta} \right)_{mnpq}~ \equiv~\frac{\d}{\d\, \eta^m}\frac{\d}{\d\, \eta^n}\frac{\d}{\d\, \eta^p}\frac{\d}{\d\, \eta^q}\ .
$$
These transformations preserve chirality, the inhomogeneous $\delta_{\bf 70}^{(-1)}\,\varphi$ because of its global character, that is 
$\d^{+}\overline\beta_{ijkl} ~=~0$, while $\delta_{\bf 70}^{(1)}\,\varphi$  is manisfestly chiral because of its coherent state-like construction.

Consider the embedding 

\be
E_8\supset SO(16)\ ,  \qquad
{\bf 248} ~=~ \bf 120\,+\, \bf128 \ ,
\ee
where the $SO(16)$ irreducible representations are decomposed in terms of $SU(8)\times U(1)$ as 

\begin{eqnarray}\label{e8toso16}
\bf 120&=&\bf63^{}_{0}\,+\,\bf28^{}_{-1}\,+\,\overline{\bf 28}^{}_{1}\,+\,\bf1^{}_0 \nn\\
\bf128 &= &\bf1_{2}'\,+\, \bf28'_{1} \,+\,\bf70_{0}\,+\,\overline{\bf28}'_{-1} \,+\,\bf\bar1'_{-2}\ ,
\end{eqnarray}
where the subscript indicates their $U(1)$ values. We recognize the $\bf 70$ as the representation in  $\Ese/SU(8)$; the rest of the coset $\Eei/SO(16)$ transformations form two $U(1)$ singlets, a twenty-eight dimensional representation and its complex conjugate (not to be confused with the $\bf 1$,  $\bf 28$, and $\bf \overline{28}$ in the adjoint representation of $SO(16)$). Closure of the algebra

$$ [ \, SO(16) \, ,\, \Eei/SO(16)\,]\, ~\subset ~\Eei/SO(16) \ ,$$
enables us to determine the $\Eei/SO(16)$ action on the chiral superfield. The construction of the $128$ inhomogeneous transformations  begins with the commutator   

$$ \delta^{(-1)}_{\bf 28'}\, \varphi~=~[\, \delta_{\bf 28}\,,\, \delta^{(-1)}_{\bf 1'} \,]\,\varphi ~=~\delta_{\bf 28}\,\delta^{(-1)}_{\bf 1'} \,\varphi- \delta^{(-1)}_{\bf 1'} \,\delta_{\bf 28}\,\varphi\ .$$
Since  the variations act only on the superfield, and $\delta^{(-1)}_{\bf 1'}\,\varphi$ is a constant which is not to be varied, this requirement amounts to 
 expressing  $\delta^{(-1)}_{\bf 28'} \, \varphi$  in terms of $\delta^{(-1)}_{\bf 1'} \, \varphi$
 
\be
\delta^{(-1)}_{\bf 28'} \, \varphi
~=~-\, \alpha_{ij}\,\frac{q^i\,q^j}{\d^+}\, \delta^{(-1)}_{\bf 1'}\,\varphi
~=~ 2\,\alpha_{ij}\,\theta^i\,\theta^j\,\d^+\, \delta^{(-1)}_{\bf 1'}\,\varphi\ ,
\ee
using    

\begin{equation}\label{q-turns-into-theta}
q^m\, \varphi(y)~=~  i\sqrt{2}\,\theta^m\,\d^+\, \varphi(y)\ .
\end{equation}
Proceeding in a similar fashion, the remaining $\Eei/SO(16)$ inhomogeneous transformations are found to be 

\begin{eqnarray}
 \delta^{(-1)}_{\bf 70} \, \varphi&=&[\, \delta_{\bf 28}\,,\, \delta^{(-1)}_{\bf 28'} \,]\,\varphi 
~\sim~ \theta^{ijkl}\,\d^{+2}\, \delta^{(-1)}_{\bf 1'}\,\varphi\ ,\nn\\
 \delta^{(-1)}_{\bf \overline{28}'} \, \varphi&=&\,[\, \delta_{\bf 28}\,,\, \delta^{(-1)}_{\bf 70} \,]\,\varphi ~\sim~\theta^{ijklmn}\,\d^{+3}\, \delta^{(-1)}_{\bf 1'}\,\varphi\ ,\nn\\
 \delta^{(-1)}_{\bf \overline{1}'} \, \varphi&=&\,[\, \delta_{\bf 28}\,,\, \delta^{(-1)}_{\bf \overline{28}'} \,]\,\varphi ~\sim~\theta^{ijklmnpq}\,\d^{+4}\, \delta^{(-1)}_{\bf 1'}\,\varphi\ .\nn
\end{eqnarray}
It is convenient to express the inhomogeneous transformations for  the $128$ parameters $\bar\beta$, $\bar\beta_{ij}$, $\bar\beta_{ijkl }$, $\bar\beta_{ijklmn}$, and $\bar\beta_{ijklmnpq}$, on the superfield $\hat\varphi=\frac{1}{\d^{+2}}\varphi$, starting with  

$$ \delta^{(-1)}_{\bf 1'} \, \hat \varphi(y)~=~ \frac{1}{\d^{+4}}\, \delta^{(-1)}_{\bf 1'} h(y)~=~ \frac{1}{\ka}\,\frac{1}{\d^{+4}} \beta(y)\ ,
$$
together with

\begin{eqnarray}
\delta^{(-1)}_{\bf 28'} \, \hat\varphi&=&i\,\frac{1}{\ka}\,\,\theta^{ij}\, \frac{1}{\d^{+3}} \overline\beta_{ij} \ ,\qquad 
 \delta^{(-1)}_{\bf 70'} \, \hat\varphi~=~-\frac{1}{\ka}\,\,\theta^{ijkl}\,\frac{1}{\d^{+2}} \overline\beta_{ijkl} ,\label{inhom70}\nn\\
 \delta^{(-1)}_{\bf \overline{28}'} \, \hat\varphi&=&i\,\frac{1}{\ka}\,\,\theta^{ijklmn}\,\frac{1}{\d^{+}}\overline\beta_{ijklmn} ,\qquad
 \delta^{(-1)}_{\bf \overline{1}'} \,\hat \varphi ~=~4\,\frac{1}{\ka}\,\,\theta^{ijklmnpq}\,\,\overline\beta_{ijklmnpq} \ .\label{inhom1}\nn
\end{eqnarray}
In this way we need only consider the operation of $\frac{1}{\d^{+n}}$ on a constant function of the chiral coordinates; it is defined in terms of integrals over the chiral coordinate 

$$\frac{1}{\d^{+n}} c(y) ~=~ (-)^n\,\frac{y^{-n}}{n!}\,c_n\,  +  \,  (-)^{(n-1)}\,\frac{y^{-(n-1)}}{(n-1)!}\,c_{n-1}\,+\, \cdots\,-y^{-} c_1\, +\, c_0\ ,$$
where $c_n$ are the integration constants. 

On the component fields these correspond to constant shifts on the boson fields only

\begin{eqnarray}
 \delta^{(-1)}_{\bf {1}'} \, h(y) ~=~\frac{\beta}{\ka} \, &\quad& \delta^{(-1)}_{\bf \overline{1}'} \, \bar h(y) ~=~\frac{\overline\beta}{\ka}\,
\nn\\
\delta^{(-1)}_{\bf 28'} \,\overline{B}_{ij}(y) ~=~  \frac{\overline \beta_{ij}}{\ka}\ ,
&\quad & \delta^{(-1)}_{\bf \overline{28}'} \, {B}^{ij}(y) ~=~\frac{\beta^{ij}}{\ka}\ ,\label{inhom28}\nn\\
 \delta^{(-1)}_{\bf 70'} \,  \overline{D}_{ijkl}(y) &=& \frac{\overline\beta_{ijkl}}{\ka}\ ,\label{inhom70comp}\nn
 \end{eqnarray}
where 

$$\overline\beta=\epsilon^{ijklmnpq}_{}\overline\beta_{\,ijklmnpq}^{}/8!\ ,\qquad \beta^{ij}_{}~=~\epsilon_{}^{ijklmnpq}\overline \beta^{}_{klmnpq}/6!\ . $$
Having determined the inhomogeneous transformations, we use a similar method to find the order-$\kappa$ coset transformations, starting  from the commutator

$$  \delta_{\bf \overline{28}'}\, \varphi~=~[\, \delta_{\bf 28}\,,\,\delta_{\bf 70} \,]\, \varphi ~=~ \delta_{\bf 28}\delta_{\bf 70}\varphi\,-\,\delta_{\bf 70}\delta_{\bf 28}\varphi  \ . $$
The symmetry  under the interchange of $\eta$ and $ -\,\eta$ in the coherent state-like form of $\delta^{(1)}_{\bf 70}\varphi$,  eq.(\ref{70transf}) then leads to

$$\,\delta_{\bf 28}\delta^{(1)}_{\bf 70}\varphi ~=~ -\,2 \,\alpha_{ij}\, \frac{\ka}{4!}\,\beta^{\,mnpq}\left(\frac{\d}{\d\, \eta} \right)_{mnpq}\,\frac{1}{\d^{+2}} \left( e^{\eta\cdot \hat{\bar d} }  \,\theta^i\theta^j\,\d^{+4}\varphi\, e^{-\eta\cdot \hat{\bar d} } \d^{+3}\varphi\,\right)\bigg|_{\eta=0}\ ,
$$
and

\begin{equation}
\delta^{(1)}_{\bf 70}\delta_{\bf 28}\varphi 
~=~\frac{-\,2\,\ka}{4!}\,\alpha_{ij}\, \theta^i\theta^j \,\beta^{\,mnpq}\left(\frac{\d}{\d\, \eta} \right)_{mnpq}\,\frac{1}{\d^{+2}} \left( e^{\eta\cdot \hat{\bar d} }  \,\d^{+4}\varphi\, e^{-\eta\cdot \hat{\bar d} } \d^{+3}\varphi\,\right)\bigg|_{\eta=0}\ .\nn
\end{equation}
Using 

$$ [\, e^{\eta\cdot \hat{\bar d} }\,,\, \theta^i \,]~=~ \frac{\eta^i}{\d^+} e^{\epsilon\cdot \hat{\bar d} }\ ,$$
one rewrites $\delta^{(1)}_{\bf 70}\delta^{}_{\bf 28}\varphi$ as

$$
-\,2 \,\alpha_{ij}\, \frac{\ka}{4!}\,\beta^{\,mnpq}\left(\frac{\d}{\d\, \eta} \right)_{mnpq}\,\frac{1}{\d^{+2}} \left( \,\left[ 
\eta^i \eta^j +\theta^i\theta^j\d^{+2}
\right] e^{\eta\cdot \hat{\bar d} } \d^{+2}\varphi\, e^{-\eta\cdot \hat{\bar d} } \d^{+3}\varphi\,\right)\bigg|_{\eta=0}\  ,
$$
which yields

\be
\delta^{(1)}_{\bf \overline{28}'}\, \varphi ~=~
 - \,2 \,\alpha_{ij}\, \frac{\ka}{4!}\,\beta^{\,mnpq}\left(\frac{\d}{\d\, \eta} \right)_{mnpq}\,\eta^i \eta^j \,\frac{1}{\d^{+2}} \left( \, e^{\eta\cdot \hat{\bar d} } \d^{+2}\varphi\, e^{-\eta\cdot \hat{\bar d} } \d^{+3}\varphi\,\right)\bigg|_{\eta=0}\ ,\nn
\ee
and can be rewritten as 
 
\be
\delta^{(1)}_{\bf \overline{28}'}\, \varphi~=~\ka\,\beta^{ij}\,  \left(\frac{\d}{\d\,\eta} \right)_{ij}\,\frac{1}{\d^{+}} \left( \, e^{\eta\cdot \hat{\bar d} } \d^{+2}\varphi\, e^{-\eta\cdot \hat{\bar d} } \d^{+2}\varphi\,\right)\bigg|_{\eta=0}\ , 
\ee
where we have reset the parameters to 

\[
\beta^{\,ij}~=~-\,\frac{1}{2}\,\beta^{\,ijmn}\,\alpha_{mn}\ .
\]
The remaining  order-$\ka$ coset transformations follow:

\begin{eqnarray}
\delta^{(1)}_{\bf 28'}\,\varphi&=&[ \, \delta_{\bf \overline{28}}\,,\, \delta^{(1)}_{\bf 70} \,]\,\nn\\
&=&{\ka}\,\beta^{\,ijmnpq}\left(\frac{\d}{\d\, \eta} \right)_{ijmnpq} \,\frac{1}{\d^{+3}} \left( \,  e^{\eta\cdot \hat{\bar d} } \d^{+4}\varphi\, e^{-\eta\cdot \hat{\bar d} } \d^{+4}\varphi\,\right)\bigg|_{\eta=0\,}\ ,\nn\\
\delta^{(1)}_{\bf 1'}\,\varphi &=&[ \, \delta_{\bf \overline{28}}\,,\, \delta^{(1)}_{\bf {28}'}\,]\,\varphi \nn\\
&=&{\ka}\,\beta^{\,ijklmnpq}\left(\frac{\d}{\d\, \eta} \right)_{ijklmnpq} \,\frac{1}{\d^{+4}} \left( \, e^{\eta\cdot \hat{\bar d} } \d^{+5}\varphi\, e^{-\eta\cdot \hat{\bar d} } \d^{+5}\varphi\,\right)\bigg|_{\eta=0}\ ,
\quad \nn\\
\delta^{(1)}_{\bf \bar 1'}\,\varphi&=&[ \, \delta^{}_{\bf 28}\,,\, \delta^{(1)}_{\bf \overline{28}'}\,]\,\varphi 
~=~ 4\,\ka \overline\beta \, \d^+\varphi \,\d^+\varphi \  .\nn
\end{eqnarray}
All these $\Eei/SO(16)$ coset transformations can be written in the compact form

\bea
&&\delta^{}_{\Eei/SO(16)}\,\varphi~=~\frac{1}{\kappa}\,F\,+\,\ka\,\ep^{i_1i_2\dots i_8}\,\sum_{c=-2}^{2}
\left(\frac{1}{i^{|c+2|}}\,\hat{\overline d}_{i_1i_2\cdots i_{2(c+2)}} \partial^{+(4+c)}_{}\,F\right)\bigg|_{\bar\theta=0}\\
&&\quad\times
\Bigg\{ \left(\frac{\d}{\d\, \eta} \right)_{i_{2c+5}\cdots i_8}\,\d^{+(c-2)} \left( e^{\eta\cdot \hat{\bar d} }  \,\d^{+(3-c)}\varphi\, e^{-\eta\cdot \hat{\bar d} } \d^{+(3-c)}\varphi\,\right)\bigg|_{\eta=0}
\,+\, \m O(\ka^2)\Bigg\}\nn\ ,
\eea
where the sum is over the $U(1)$ charges $c=2,1,0,-1,-2$ of the bosonic fields, and

\begin{eqnarray}
F&=&\,\frac{1}{{\d^+}^2}\,\beta\,(y)\,\,+\,i\,\theta^{mn}_{}\,\frac{1}{\d^+}\,{\overline \beta}_{mn}\,(y)-\,\theta^{mnpq}_{}\,{\overline \beta}^{}_{mnpq}\,(y)+\nn \\
&&+\,i\widetilde\theta^{}_{~mn}\,\d^+\,\beta^{mn}\,(y)+\,{4}\,\widetilde\theta\,{\d^+}^2\,{\bar \beta}\,(y)\ ,\nn
\end{eqnarray}
and

$$\hat{\overline d}_{i_1i_2\cdots i_{2(c+2)}} ~\equiv~ \frac{1}{(2c+4)!}\, \hat{\overline d}_{i_1}\hat{\overline d}_{i_2}\cdots\hat{\overline d}_{2(c+2)}\ .$$ 

This construction can in principle be continued to order $\kappa^3$, but its expression would not yield any further insight. It is to be emphasized that this symmetry is independent of dynamics. However supersymmetric dynamics in various dimensions may or may not respect it. In $d=3$, as we show in the next section, it is left intact, but it is progressively nibbled at in higher dimensions, until nothing is left of it in $d=11$.

%%%%%%%%%%%%%%%%%%%%%%%%%%%%%%%%%%%%%%%%%%%%%%%%%%
\section{$\Eei$-Invariant Dynamics}
In supersymmetric theories, the Hamiltonian is determined from the dynamical supersymmetries. Thus its invariance under any symmetry requires the dynamical supersymmetries to have well-defined transformation properties. 

Invariance of the Hamiltonian under $\Eei$ requires the dynamical supersymmetries to transform linearly under $SO(16)$. It is easy to see that this restricts the dynamics to take place in three space-time dimensions. In four dimensions, the lowest order dynamical supersymmetries, with parameters  $\ep^m$ and $\bar \ep_m$, 

\begin{eqnarray}
\delta^{dyn}_s\,\varphi~=~\epsilon^m \frac{\partial}{\partial^+}\,\bar q_m\,\varphi\,+\, \m O(\kappa)\ ,&&
\delta^{dyn}_{\bar s}\,\varphi~=~\bar\epsilon_m \frac{\bar \partial}{\partial^+}\, q^m\,\varphi\,+\, \m O(\kappa)\ ,
\end{eqnarray}
transform under $SU(8)$ as $\bf \bar 8$  and $\bf 8$, respectively; they lead to $\Ese$-invariant dynamics. It is easy to see that they do not transform into one another under  $SO(16)/SU(8)$ unless the transverse derivatives satisfy $\d=\bar \d$. This is automatic in $d=3$ where there is only one transverse space dimension; only then the dynamical supersymmetries transform as the vectorial $\bf 16$ of $SO(16)$, and we are dealing with a theory with $\m N=16$ supersymmetries.

In order to construct the dynamical supersymmetries to higher orders in $\kappa$, we note that although there is no helicity in three dimensions, $SO(16)$ requires covariance for its  $U(1)$ subgroup. Its action on the superfield, eq.(\ref{SU8andU1}),

$$ \delta_{U(1)}\, \varphi ~=~ T\, \varphi~=~\Big(\, 2 \, -\,\frac{1}{2} \theta^m\bar q_m\,\Big)\, \varphi \ , $$
is rewritten here in terms of linear operators,  assigning  the charge $+2$ to $\varphi$. Each component of the superfield has a definite $U(1)$ value:  $h$ has value $+2$
 
\be
\delta_{U(1)}h ~=~ 2 h\ ,
\ee
and the $U(1)$ charges of the remaining bosons are $\overline{B}_{ij}(1)$, $\overline{D}_{ijkl}(0)$, ${B}^{ij}(-1)$, $\overline h(-2)$, and for the fermions $\overline{\psi}_i(3/2)$,
$\overline{\chi}_{ijk}(1/2)$, ${\chi}^{ijk}(-1/2)$, and ${\psi}^i(-3/2)$. It follows that the dynamical supersymmetry transformation has a definite charge, that is  

\be
[\, \delta^{}_{U(1)}\,,\delta^{dyn}_{s}\,] \,\varphi ~=~-\frac{1}{2}\, \delta^{dyn}_{s}\,\varphi \ .
\ee
Any term in $\delta^{dyn}_{s}\,\varphi$ which is of higher order in the superfields must have the same charge as the linear term. 

This is not possible for the quadratic term: using the inside-out constraint, the charge of $\bar\varphi$ is opposite that of $\varphi$, so either we have $\varphi\varphi$ with twice the charge,  or $\varphi\overline\varphi$ with no charge; either way neither has the same  charge as that of the term linear in $\varphi$. We conclude that the dynamical supersymmetries contain no terms linear in $\kappa$: the Hamiltonian has no cubic interaction. 

The same is not true for the order $\kappa^2$ term cubic in the superfield; there we can have terms structurally of the form 

\begin{equation}\label{structurally}
 \quad\ep^m \bar q_m \varphi\, \varphi\, \bar d^8 \varphi \ \sim\ 
\ep^m \bar q_m \varphi\, \varphi\, \overline\varphi\  ,
\end{equation}
by which we mean three chiral superfields with eight powers of $\bar d$ sprinkled among them. The quartic interaction in the Hamiltonian, constructed from the free and order $\ka^2$ dynamical supersymmetries with the same charge, can now be $U(1)$ invariant.

The two supersymmetries are obtained from one another by 

\begin{equation}
[\, \delta^{}_{\bf 28}\,,\delta^{dyn}_{s}\,] \,\varphi ~=~ \delta^{dyn}_{\bar s}\,\varphi \ ,
\end{equation}
which must be true to all orders in $\kappa$. As we did in \cite{Brink:2008qc}, the dynamical supersymmetry transformations are restricted by requiring that they commute with the non-linear part of the symmetry, that is 

$$ [ \, \delta^{}_{\Eei/SO(16)}\,,\,\delta^{dyn\,}_{s} \, ] \, \varphi ~=~0\ .$$
Expanding this equation in the coupling shows that $\delta^{dyn\,}_{s}\varphi$ contains only terms with oven powers of $\kappa$, since $U(1)$ invariance forbids terms quadratic in the superfield.  To first order in $\ka$, we find that

$$ [ \, \delta^{(1)}_{\Eei/SO(16)}\,,\,\delta^{dyn\,(0)}_{s} \, ] \, \varphi\,+\, [ \, \delta^{(-1)}_{\Eei/SO(16)}\,,\,\delta^{dyn\,(2)}_{s} \, ] \, \varphi ~=~0\ ,$$ 
which is used to restrict the form of $\delta^{dyn\,(2)}_{s}\varphi$. Coupled with the $U(1)$ charge restriction ($\delta^{dyn\,(2)}_{s}\,\varphi$  built out of three chiral superfields with eight $\bar d_m$'s), this equation is sufficient to determine its form.

To see this, choose a particular  $\Eei/SO(16)$ transformation, say $\bf\bar 1'$, which yields   

$$
\delta^{(-1)}_{\bf \bar 1' } \delta^{dyn\,(2)}_{s}\,\varphi
~=~\bar \beta\frac{\ka}{\d^+} \left( \ep\bar q \d \varphi\, \d^{+2}\varphi\,-\,\d \d^+\varphi \, \ep\bar q \d^+ \varphi\right)\ , $$
since $\delta^{(-1)}_{\bf \bar 1' }\,\varphi$ is a constant, and therefore $\delta^{dyn\,(2)}_{s}\delta^{(-1)}_{\bf \bar 1' }\,\varphi$ vanishes. It constrains only the terms in $\delta^{dyn\,(2)}_{s}\varphi$ which are affected by $\delta^{(-1)}_{\bf \bar 1' }$, which we denote by $\delta^{dyn\,(2)\,\,[{\bf 1'}]}_{s}\,\varphi$. 

By introducing the  operators 

$$E ~\equiv~ e^{a\hat{\d}\,+\, b\, \ep\hat{\bar q}\,+\, \eta\hat{d}}\, ~{\rm and}~E^{-1} ~\equiv~ e^{-\,a\hat{\d}\,-\, b\, \ep\hat{\bar q}\,-\, \eta\hat{d}}\  ,$$
we can rewrite this constraint in compact form 

\begin{equation}\label{boundarycond1'}
\delta^{(-1)}_{\bf \bar 1' } \delta^{dyn\,(2)}_{s}\,\varphi
~=~ \ka\,\frac{\bar\beta}{2}\,\frac{\d}{\d a}\frac{\d}{\d b}\left(\, E \d^{+2}\varphi\,\, E^{-1} \d^{+2}\varphi \,\right)\Big|_{a=b=\eta=0}\ .
\end{equation}
Consider the chiral combination

\begin{equation}\label{sol1'bar}
\delta^{dyn\,(2)\, [{\bf \bar 1'}]}_{s}\,\varphi~=~  \frac{\ka^2}{2} \frac{\d}{\d a}\frac{\d}{\d b}\frac{1}{\d^{+2}}\Big[ E\d^{+3} \varphi\, \, E^{-1} Z \Big] \,\Big|_{a=b=\eta=0}\ ,
\end{equation}
where  	

$$ Z ~=~ \frac{\ep^{ijklmnpq}}{2\cdot 8!}\left(\frac{\d}{\d\xi}\right)_{ijklmnpq}\frac{1}{\d^{+4}}\left(  e^{\xi \hat{\bar d}} \d^{+6} \varphi \,\,  e^{-\xi \hat{\bar d}} \d^{+6} \varphi\right)\,\Big|_{\xi=0} \ .$$
By taking $\delta^{(-1)}_{\bf \bar 1'}$ on this chiral combination (\ref{sol1'}), one gets

\begin{eqnarray}
\delta^{(-1)}_{\bf \bar 1'}\delta^{dyn\,(2)\, [{\bf \bar 1'}]}_{s}\,\varphi&=&
\ka^2 \frac{\bar\beta}{2} \frac{\d}{\d a}\frac{\d}{\d b}\frac{1}{\d^{+2}}\Big[ E\d^{+3} \delta^{(-1)}_{\bf \bar 1'} \varphi\, \, E^{-1} Z \,+\, E\d^{+3}  \varphi\, \, E^{-1} \delta^{(-1)}_{\bf \bar 1'} Z \Big] \,\Big|_{a=b=\eta=0}\nn\\
&=&\ka\,\frac{\bar\beta}{2}\,\frac{\d}{\d a}\frac{\d}{\d b}\frac{1}{\d^{+2}}\left(\, E \d^{+3}\varphi\,\, E^{-1} \d^{+2}\varphi \,\right)\Big|_{a=b=\eta=0}\ ,%\nn\\
\end{eqnarray}
where the first term $E\d^{+3} \delta^{(-1)}_{\bf \bar 1'} \varphi$ vanishes since $\d^{+3} \delta^{(-1)}_{\bf \bar 1'} \varphi~=~\frac{1}{\ka} \theta^8\d^+ \bar \beta~=~0$, and the second term becomes

$$\delta^{(-1)}_{\bf \bar 1'} Z ~=~\frac{\bar \beta}{\ka} \,\d^{+2} \varphi\ . $$
Thus, this chiral combination is the solution that satisfies the constraint (\ref{boundarycond1'}).

The dynamical supersymmetry transformations for  the rest of the coset $\Eei/SO(16)$, $\overline{\bf 28}'$, $\bf 70$, $\bf 28'$ and $ \bf  1'$,  can be obtained in a similar fashion.
For the $\bf \overline{28}'$ transformations, commutativity yields the constraint 
 
$$
\delta^{(-1)}_{\overline{\bf 28}'} \delta^{dyn\,(2)}_{s}\,\varphi~=~\frac{\ka}{2}
\beta^{ij}\left(\frac{\d}{\d\eta}\right)_{ij} 
\frac{\d}{\d a}\frac{\d}{\d b}
\frac{1}{\d^{+2}}\left( E\, \d^{+3} \varphi\,\, E^{-1}\, \d^{+3} \varphi
\right)\Big|_{a=b=\eta=0}\ ,
$$
and with the solution 

\begin{equation}\label{solbar28'}
\delta^{dyn\,(2)\, [{\bf \overline{28}'}]}_{s}\,\varphi~=~  \frac{\ka^2}{2} \left(\frac{\d}{\d\eta}\right)_{ij} \frac{\d}{\d a}\frac{\d}{\d b}\frac{1}{\d^{+3}}\Big[ E\d^{+4} \varphi\, \, E^{-1} Z^{ij} \Big] \,\Big|_{a=b=\eta=0}\ ,
\end{equation}
where $Z^{ij}$ is defined as 	

$$ Z^{ij} ~=~  \frac{\ep^{ijklmnpq}}{2\cdot 6!}\left(\frac{\d}{\d\xi}\right)_{klmnpq}\frac{1}{\d^{+2}}\left(  e^{\xi \hat{\bar d}} \d^{+5} \varphi \,\,  e^{-\xi \hat{\bar d}} \d^{+5} \varphi\right)\,\Big|_{\xi=0} \ .$$

The constraint for the $\bf {70}$ transformations is 

$$
\delta^{(-1)}_{\bf {70}} \delta^{dyn\,(2)}_{s}\,\varphi~=~\frac{\ka}{2}
\beta^{ijkl}\left(\frac{\d}{\d\eta}\right)_{ijkl} 
\frac{\d}{\d a}\frac{\d}{\d b}
\frac{1}{\d^{+3}}\left( E\, \d^{+4} \varphi\,\, E^{-1}\, \d^{+4} \varphi
\right)\Big|_{a=b=\eta=0}\ ,
$$
which is the same as for $\m N=8$ Supergravity, and the solution is therefore of the same form

\begin{equation}\label{sol70}
	\delta^{dyn\,(2)\, [{\bf {70}}]}_{s}\,\varphi~=~  \frac{\ka^2}{2} \left(\frac{\d}{\d\eta}\right)_{ijkl} \frac{\d}{\d a}\frac{\d}{\d b}\frac{1}{\d^{+4}}\Big[ E\d^{+5} \varphi\, \, E^{-1} Z^{ijkl} \Big] \,\Big|_{a=b=\eta=0}\ ,
\end{equation}
where  	

$$ Z^{ijkl} ~=~  \frac{\ep^{ijklmnpq}}{2\cdot 4!}\left(\frac{\d}{\d\xi}\right)_{mnpq}\left(  e^{\xi \hat{\bar d}} \d^{+4} \varphi \,\,  e^{-\xi \hat{\bar d}} \d^{+4} \varphi\right)\,\Big|_{\xi=0} \ .$$
Repeating the same procedure, one obtains the constraints from  the $\bf 28'$ transformations 

$$
\delta^{(-1)}_{\bf {28'}} \delta^{dyn\,(2)}_{s}\,\varphi~=~\frac{\ka}{2}
\beta^{ijklmn}\left(\frac{\d}{\d\eta}\right)_{ijklmn} 
\frac{\d}{\d a}\frac{\d}{\d b}
\frac{1}{\d^{+4}}\left( E\, \d^{+5} \varphi\,\, E^{-1}\, \d^{+5} \varphi
\right)\Big|_{a=b=\eta=0}\ ,
$$
with the solution

\begin{equation}\label{sol28'}
\delta^{dyn\,(2)\, [{\bf {28}'}]}_{s}\,\varphi~=~  \frac{\ka^2}{2} \left(\frac{\d}{\d\eta}\right)_{ijklmn} \frac{\d}{\d a}\frac{\d}{\d b}\frac{1}{\d^{+5}}\Big[ E\d^{+6} \varphi\, \, E^{-1} Z^{ijklmn} \Big] \,\Big|_{a=b=\eta=0}\ ,
\end{equation}
where 	

$$ Z^{ijklmn} ~=~  \frac{\ep^{ijklmnpq}}{2\cdot 2!}\left(\frac{\d}{\d\xi}\right)_{pq}\d^{+2}\left(  e^{\xi \hat{\bar d}} \d^{+3} \varphi \,\,  e^{-\xi \hat{\bar d}} \d^{+3} \varphi\right)\,\Big|_{\xi=0} \ .$$
Finally, the  $\bf 1'$ transformations yield

$$
\delta^{(-1)}_{\bf {1'}} \delta^{dyn\,(2)}_{s}\,\varphi~=~\frac{\ka}{2}
\beta^{ijklmnrs}\left(\frac{\d}{\d\eta}\right)_{ijklmnrs} 
\frac{\d}{\d a}\frac{\d}{\d b}
\frac{1}{\d^{+5}}\left( E\, \d^{+6} \varphi\,\, E^{-1}\, \d^{+6} \varphi
\right)\Big|_{a=b=\eta=0}\ ,
$$
together with the solution
\begin{equation}\label{sol1'}
\delta^{dyn\,(2)\, [{\bf {1}'}]}_{s}\,\varphi~=~  \frac{\ka^2}{2} \left(\frac{\d}{\d\eta}\right)_{ijklmnpq} \frac{\d}{\d a}\frac{\d}{\d b}\frac{1}{\d^{+6}}\Big[ E\d^{+7} \varphi\, \, E^{-1} Z^{ijklmnpa} \Big] \,\Big|_{a=b=\eta=0}\ ,
\end{equation}
where 	

$$ Z^{ijklmnpq} ~=~  \frac{1}{2}\ep^{ijklmnpq} \d^{+4}\left(  e^{\xi \hat{\bar d}} \d^{+2} \varphi \,\,  e^{-\xi \hat{\bar d}} \d^{+2} \varphi\right)\,\Big|_{\xi=0} \ .$$

Combining all together, one writes a compact form for the constraints as a sum over the five $U(1)$ values of the coset  transformations

\begin{eqnarray}\label{order-kappa}
&&\delta^{(-1)}_{\Eei/SO(16)} \delta^{dyn\,(2)}_{s}\,\varphi \\
&&=\frac{\ka}{2}
\sum^{2}_{c=-2} \beta^{i_1\cdots i_{2(2+c)}} \left(\frac{\d}{\d\eta}\right)_{i_1\cdots i_{2(2+c)}} 
%\\
%&&
\frac{\d}{\d a}\frac{\d}{\d b}\frac{1}{\d^{+(3+c)}}\left( E\, \d^{+(4+c)} \varphi\,\, E^{-1}\, \d^{+(4+c)} \varphi
\right)\Big|_{a=b=\eta=0}\ . \nn
\end{eqnarray}
Their solutions are given by

\begin{eqnarray}\label{final}
&&\delta^{dyn\,(2)}_{s}\,\varphi \nn\\
&&=~\frac{\kappa^2}{2}\, \sum^{2}_{c\,=\,-2}\frac{1}{\d^{+(c+4)}} \Bigg\{ 
\frac{\d}{\d a}\frac{\d}{\d b} \left(\frac{\d}{\d\eta}\right)_{i_1i_2\cdots i_{2(c+2)}}\left(E \d^{+(c+5)} \varphi E^{-1}\right)\Bigg|_{a=b=\eta=0}\nn
\\
&&~  \times \,\frac{\ep^{i_1i_2\cdots \i_{i_8}}}{(4 -2\,c)!}
 \left(\frac{\d}{\d\, \eta} \right)_{i_{2c+5}\cdots i_8}\d^{+2c} \left(
E  \d^{+(4-c)} \varphi E^{-1}  \d^{+(4-c)} \varphi 
 \right)\Bigg|_{\eta=0}\Bigg\}\ ,~~
\end{eqnarray}
where the sum is, as before, over the $U(1)$ charges.  One term in (4.14), the one from the variation generated by the $\bf 70$, eq.(4.10), is in fact the expression we get at this order by dimensionally reducing the $d=4$ theory.

The absence of a term of order-$\kappa^{2n+1}$ in the dynamical supersymmetry transformations means that the Hamiltonian itself has no order-$\kappa^{2n+1}$ interactions. This is not a surprise since the chiral superfield contains the two spinor representations of $SO(16)$, $\bf 128$ and $\bf 128'$, and spinor representations have no odd-order invariants. This shows that the $\Eei$-invariant theory is distinct from the other $\m N=16$ supergravity theory  that is obtained by dimensional reduction from $\m N=8$ Supergravity \cite{Brink:2008qc} in four dimensions. 

The dynamical supercharge is the basic construct. The Hamiltonian is easily obtained by either using the anticommutator between the dynamical supercharge and its complex conjugate, or from the using the quadratic form as in \cite{Brink:2008qc}. We do not do it here even though it is straightforward since it does not add to our knowledge about this theory.

%%%%%%%%%%%%%%%%%%%%%%%%%%%%%%%%%%%%%%%%%%%%%%%%%%%
\section{Conclusions and Outlook}
In this paper we have shown how to construct the  $\Eei$ symmetry on the maximally supersymmetric light-cone superfield with $256$ degrees of freedom. The complete symmetry spanned by this superfield  is a semi-direct product of the superPoincar\'e symmetry and the  $\Eei$ symmetry. This sounds somewhat strange since there is no such supersymmetry in classifications of superalgebras. The key point here is that when the  $\Eei$ symmetry is decomposed into $SO(16) \times  \Eei /SO(16) $, the superalgebra transforms under the $SO(16)$ but not under the coset. This is possible only because the coset is non-linearly realized. This gives us then a powerful method to construct the dynamics where the coset transformations can be used order by order to find the dynamics.

It is clear from the construction that this is not a priori a dimensionally reduced theory from the  $d=4$ one. If the dynamical supersymmetry derived in (\ref{final}) is ``oxidized" to $d=4$ it will not transform correctly under the helicity generator. It is usually argued that when the maximal supergravity is dimensionally reduced to $d=3$ one has to use duality transformations  and Weyl scalings to get all the bosonic fields to be scalars, and it is only then that one can find the $\Eei$ symmetry.  In the light-cone formulation where only physical degrees of freedom are present, the duality of a vector with a scalar in $d=3$ is trivial in the sense that it reduces to an identity. The one dynamical component of a vector field does indeed transform as a scalar. We have been unable to find a field redefinition, which is the only freedom we can try here, to connect the seemingly different two $d=3$ theories: the one with the full $\Eei$ symmetry constructed here and the other with $E_{7(7)}$ obtained by a naive dimensional reduction.

The complete one-loop contribution to the four-graviton scattering matrix element in any dimension  was constructed in \cite{Green:1982sw} using the zero-slope limit and dimensional reduction of Type II superstrings.  It was found to be a box diagram with the proper kinematical factors. Naively such result can only be derived from an underlying field theory with a three-point coupling.  This can be seen by looking at the cuts of the amplitude. However there is no three-point coupling in the theory with an  $\Eei$ symmetry as we have discussed above and thus we conclude that the $d=3$ theory that we have derived above would give a different one-loop contribution to the four-graviton scattering matrix element. We note, however, that in $d=3$ the infrared singularities are worse than in higher dimensions and the loop amplitude has to be very carefully constructed. That might resolve the problem of relating the two theories but we have again been unable to do so. The $d=3$ theory that we have constructed is unique and must be the one constructed by Marcus and Schwarz \cite{Marcus:1983hb} and by de Wit, Nicolai and Tollsten \cite{de Wit:1992up}.

There are now two ways to continue this analysis. We could try to go down in dimensions and in that process try to find the infinite algebras $E_9$, $E_{10}$ and possibly $E_{11}$. These algebras have been quite popular recently with claims that they play a r\^ole also for the higher-dimensional theories.  For a review, see \cite{Henneaux:2007ej} and references contained therein. The other road is to go up in dimensions and check how much is left of the exceptional symmetries in various dimensions. Our analysis suggests that the exceptional symmetries could be broken in higher dimensions in a  controlled way such that they still play a r\^ole for the dynamics. We believe that our formalism is quite suitable for the study of both these lines. 

We finally like to point out that our formalism using coherent-state techniques is extremely efficient. The expression in (\ref{final}) will contain hundreds of terms of different combinations of the superfield. With the new technique they can be treated all in one go. This gives us hope that we should be able to find expressions to all orders in the coupling constants just as in a non-linear $\sigma$-model. After all the $d=3$ theory is a supersymmetric version of a non-linear $\sigma$-model. We also hope to be able to come back this issue in future publications.

%%%%%%%%%%%%%%%%%%%%%%%%%%%%%%%%%%%%%%%%%%%%%%%%%%%

\begin{flushleft}  
{\Large \bf Acknowledgements}
\end{flushleft}
\noindent We thank Sudarshan Ananth for pointing out references on $\Eei$ invariant theories and 
Martin Cederwall and  Marc Henneaux for useful discussions. SK thanks KIAS for its hospitality during his visit where revision of the paper is made. He is supported by a McLaughlin Dissertation Fellowship from the University of Florida. 
Two of us (PR and SK) are also supported by the Department of Energy Grant No. DE-FG02-97ER41029.

%%%%%%%%%%%%%%%%%%%%%%%%%%%%%%%%%%%%%%%%%%%%%%%%%%%

\end{document}